\documentclass[12pt]{article}

\usepackage{etex}
\usepackage{amsmath,amssymb,amsthm,amscd,amsbsy,array}
\usepackage[english]{babel}

\usepackage{soul}

\usepackage{color}
\usepackage[colorlinks=true, pdfstartview=FitV, linkcolor=blue, citecolor=blue, urlcolor=blue]{hyperref}

\let\ssection=\section
\renewcommand{\section}{\setcounter{equation}{0}\ssection}

\newcommand{\bR}{\mathbb{R}}

\newcommand{\bZ}{\mathbb{Z}}

\newcommand{\arccot}{\mathop{\mathrm{arccot}}}

\newcommand{\bb}{\mathbf{b}}

\newcommand{\bbeta}{\boldsymbol{\beta}}

\newcommand{\bc}{\mathbf{c}}
\newcommand{\chr}{\mathfrak{chr}}
\newcommand{\Chr}{\mathrm{Chr}}
\newcommand{\bgamma}{\boldsymbol{\gamma}}

\newcommand{\Conf}{\mathrm{Conf}}
\newcommand{\const}{\mathrm{const.}}

\newcommand{\Div}{\boldsymbol{\nabla}\cdot}

\newcommand{\hGamma}{{\hat{\Gamma}}}
\newcommand{\bGamma}{\boldsymbol{\Gamma}}
\newcommand{\GL}{\mathrm{GL}}
\newcommand{\gl}{{\mathfrak{gl}}}
\newcommand{\bg}{{\mathbf{g}}}

\newcommand{\rg}{\mathrm{g}}

\newcommand{\cM}{\mathcal{M}}

\newcommand{\tPhi}{\varphi}

\newcommand{\Rot}{\boldsymbol{\nabla}{\times}\,}
\newcommand{\Ric}{\mathrm{Ric}}

\newcommand{\hs}{{\hat{s}}}
\newcommand{\cS}{{\mathcal{S}}}

\newcommand{\SL}{\mathrm{SL}}
\newcommand{\Sl}{\mathfrak{sl}}
\newcommand{\SO}{\mathrm{SO}}

\newcommand{\SN}{Schr\"odinger-Newton\ }

\newcommand{\sor}{\mathfrak{so}}
\newcommand{\rO}{\mathrm{O}}

\newcommand{\ts}{{\tilde{s}}}

\newcommand{\Sch}{\mathrm{Sch}}
\newcommand{\sch}{\mathfrak{sch}}

\newcommand{\wht}{{\hat{t}}}
\newcommand{\wtt}{{\tilde{t}}}

\newcommand{\SV}{\mathrm{SV}}

\newcommand{\bv}{\mathbf{v}}
\newcommand{\bx}{\mathbf{x}}

\newcommand{\hbx}{{\hat{\bx}}}
\newcommand{\tbx}{{\tilde{\bx}}}

\newcommand{\half}{\frac{1}{2}}

\newcommand{\la}{{\langle}}
\newcommand{\ra}{{\rangle}}

\def\parag{\hfil\break} %%%%% paragraph
\def\kikezd{\parag\underbar}

\newcommand{\medbox}[1]{\fbox{%
\rule[-15pt]{0pt}{35pt}$\;\;\displaystyle{#1}\;\;$}%
}

\def\beq{\begin{equation}}
\def\eeq{\end{equation}}
\def\beqa{\begin{eqnarray}}
\def\eeqa{\end{eqnarray}}

\def\barray{\left(\begin{array}}
\def\earray{\end{array}\right)}
\def\barraynb{\begin{array}}
\def\earraynb{\end{array}}

\usepackage{color}
\begin{document}

\baselineskip=17pt

\oddsidemargin .1truein
\newtheorem{thm}{Theorem}[section]
\newtheorem{lem}[thm]{Lemma}
\newtheorem{cor}[thm]{Corollary}
\newtheorem{pro}[thm]{Proposition}
\newtheorem{ex}[thm]{Example}
\newtheorem{rmk}[thm]{Remark}
\newtheorem{defi}[thm]{Definition}
%\newremark{ex}[thm]{Example}

\title{
Conformal and projective symmetries\\
in Newtonian cosmology
}

%\author{
%ChD\footnote{mailto: duval-at-cpt.univ-mrs.fr}
%+GWG\footnote{mailto: gwg1@cam.ac.uk}
%+PAH\footnote{mailto: horvathy-at-lmpt.univ-tours.fr}
%}

\author{
C. Duval\footnote{\href{mailto:duval@cpt.univ-mrs.fr}{duval@cpt.univ-mrs.fr}}
\\
{\small Centre de Physique Th\'eorique,}\\
{\small Aix Marseille Universit\'e \& Universit\'e de Toulon \& CNRS UMR 7332,}\\
{\small Case 907, 13288 Marseille, France}
\\[6pt]
G. W. Gibbons\footnote{\href{mailto:G.W.Gibbons@damtp.cam.ac.uk}{G.W.Gibbons@damtp.cam.ac.uk}}
\\
{\small D.A.M.T.P., Cambridge University,}\\
{\small Wilberforce Road, Cambridge CB3 0WA, U.K.}
\\[6pt]
P. A. Horv\'athy\footnote{\href{mailto:horvathy@lmpt.univ-tours.fr}{horvathy@lmpt.univ-tours.fr}}
\\
{\small Laboratoire de Math\'ematiques et de Physique
Th\'eorique,}\\
{\small Universit\'e de Tours, France}
\\
{\small Institute of Modern Physics, Chinese Academy of Sciences, 
Lanzhou, China}
}

\date{November 7, 2016}

\maketitle

\thispagestyle{empty}

\begin{abstract}
Definitions of non-relativistic conformal transformations are considered both in the Newton-Cartan and in the Kaluza-Klein-type Eisenhart/Bargmann geometrical frameworks. The  symmetry groups that come into play are exemplified by the cosmological, and also the Newton-Hooke solutions of Newton's gravitational field equations. It is shown, in particular, that the maximal symmetry group of the standard cosmological model is  isomorphic to the $13$-dimensional conformal-Newton-Cartan group whose conformal-Bargmann extension is explicitly worked out. Attention is drawn to the appearance of independent space and time dilations, in contrast with the Schr\"odinger group or the Conformal Galilei Algebra.
\end{abstract}

\newpage

\tableofcontents

%\newpage

%%%%%%%%%%%%%%%%%%%%%%%%%%%%%%%%%%%%%%%%%%%%%%%%%%%%%%%%%%%%%%%%%%%%%%%%%%
%%%%%%%%%%%%%%%%%%%%%%%%%%%%%%%%%%%%%%%%%%%%%%%%%%%%%%%%%%%%%%%%%%%%%%%%%%
\section{Introduction}\label{IntroductionSection}
%%%%%%%%%%%%%%%%%%%%%%%%%%%%%%%%%%%%%%%%%%%%%%%%%%%%%%%%%%%%%%%%%%%%%%%%%%
%%%%%%%%%%%%%%%%%%%%%%%%%%%%%%%%%%%%%%%%%%%%%%%%%%%%%%%%%%%%%%%%%%%%%%%%%%

One of the cardinal properties of the standard Friedmann-Lema\^{\i}tre-Robertson-Walker (FLRW) model of relativistic cosmology is its maximal conformal symmetry, arising from the vanishing of the Weyl conformal curvature. The purpose of the present article is to investigate the maximal symmetry group of an older, often unjustly dismissed, description of the non-relativistic universe at a large scale, namely Newtonian cosmology, briefly reviewed in Section \ref{NC-cosmologySection}. We found it natural to relate this search to the symmetries of general time dependent mechanical systems in the framework of Newton-Cartan (NC) geometry~\cite{Cart}.

\goodbreak

The surprising feature of a non-relativistic spacetime structure is the very specific independence of the NC connection with respect to the degenerate Galilei metric of spacetime \cite{Cart,Tra,Kun}. This requires that, unlike conformal Lorentzian sym\-metries, \textit{conformal-NC symmetries} be specified on the one hand by the pre\-servation of the space and time directions of the Galilei ``metric'', and, on the other hand, by the preservation of the projective structure associated with the symmetric NC connection ruling free fall. This procedure, spelled out in Section \ref{NC-conformalSection}, has proven critical in devising a purely geometrical definition \cite{Duv,Duv2} of the otherwise well-known \textit{Schr\"odinger group} (re)discovered in the quantum framework in the early seventies \cite{Nie,Hag} (see also \cite{DH-NC} for an  account about the (pre)history of the Schr\"odinger symmetry).

The main upshot of this article is the proof that the standard Newtonian cosmological model admits a $13$-dimensional group of (local) \textit{conformal-NC symmetries} isomorphic to the formerly named ``chronoprojective group'' \cite{Duv,BDP0,BDP} of the flat canonical NC structure (see Section %\ref{NewtonianCosmologySymmetriesSection}) 
\ref{Examples})
contains the $12$-dimensional (centreless) Schr\"odinger group as a subgroup and is \emph{different} from the usual centrally extended Schr\"odinger group.
  
We prove, \textit{en passant}, that the Cosmological Principle here is strictly equivalent to the projective flatness of the Newtonian cosmological model under investigation.

Likewise, the conformal-NC group of the model of empty space with a  cosmological constant is shown to be a $13$-dimensional which contains the Newton-Hooke group \cite{GP}, as shown in Section 
%\ref{Cosmo-NH-SymmetrySection}. 
\ref{Examples}.
The case of the vacuum static solution of the Newton field equations is treated along the same lines, highlighting the $5$-dimensional ``Virial group'' as the conformal-NC group.

In the second part of the article, namely in Section \ref{Conf-BargSymmetrySection}, we have recourse to the so-called \textit{Eisenhart-lift} \cite{Eis,Car,BM} of a NC structure. The latter allows us to deal properly with $(\bR,+)$-extensions of non-relativistic conformal symmetry groups mentioned above. This uses the full machinery of Bargmann extended spacetime structures \cite{DBKP} above NC structures.  In a nutshell, a Bargmann manifold is defined by a Lorentz metric and a null parallel vector field $\xi$, and is thus akin to a (generalized) pp-wave \cite{DGH}. 

Those structures enable us to easily redefine the notion of non-relativistic conformal symmetries as \textit{bona fide} conformal sub-symmetries of the Lorentz metric. This time, we introduce \textit{conformal-Bargmann flatness} via the vanishing of the conformal Weyl curvature, $C=(C_{\lambda\mu\nu\rho})$, subject to the condition
\begin{equation}
C_{\lambda\mu\nu\rho}\,\xi^\rho=0
\label{Cxi=0}
\end{equation}
put forward some time ago  \cite{Duv2,BDP,PBD}. The previously studied models are thus readily Eisenhart-lifted to yield their \textit{conformal-Bargmann symmetries} as \textit{non-central} extensions of their conformal-NC ones.

%%%%%%%%%%%%%%%%%%%%%%%%%%%%%%%%%%%%%%%%%%%%%%%%%%%%%%%%%%%%%%%%%%%%%%%%%%
%%%%%%%%%%%%%%%%%%%%%%%%%%%%%%%%%%%%%%%%%%%%%%%%%%%%%%%%%%%%%%%%%%%%%%%%%%
\section{Newtonian cosmology}\label{NC-cosmologySection}
%%%%%%%%%%%%%%%%%%%%%%%%%%%%%%%%%%%%%%%%%%%%%%%%%%%%%%%%%%%%%%%%%%%%%%%%%%
%%%%%%%%%%%%%%%%%%%%%%%%%%%%%%%%%%%%%%%%%%%%%%%%%%%%%%%%%%%%%%%%%%%%%%%%%%

%%%%%%%%%%%%%%%%%%%%%%%%%%%%%%%%%%%%%%%%%%%%%%%%%%%%%%%%%%%%%%%%%%%%%%%%%%
\subsection{Newton-Cartan structures}
%%%%%%%%%%%%%%%%%%%%%%%%%%%%%%%%%%%%%%%%%%%%%%%%%%%%%%%%%%%%%%%%%%%%%%%%%%

Let us first recall that the quadruple $(N,\gamma,\theta,\nabla)$ is a \textit{Newton-Cartan structure} if $N$ is a smooth $4$-dimensional manifold (spacetime), $\gamma$ a nowhere vanishing twice-contravariant tensor, $\theta$ a closed $1$-form generating the kernel of $\gamma$, and~$\nabla$ a symmetric affine connection that parallel transports $\gamma$ and $\theta$; see, e.g., \cite{Tra,Kun,DH-NC,DH}. The ``clock'', $\theta$, descends to the time axis, $T=N/\ker(\theta)$. As to the tensor $\gamma$, it gives rise to a Riemannian metric on the fibres of the projection $M\to{}T$, that is on all copies of space at fixed time.\footnote{In this Section devoted to cosmology, we restrict consideration to $4$-dimensional NC structures; almost all geometric definitions and results can be easily generalized to any spacetime dimension $d+1$.} Following~\cite{Kun}, we call such a triple $(N,\gamma,\theta)$ a \textit{Galilei structure}.

For our present purpose, $N$ will be an open subset of $\bR^4$, and 
$
\gamma=\delta^{ij}\,\partial_i\otimes\partial_j
$
and
$\theta=dt
$
where $t=x^4$ is the absolute time, in a Galilean coordinate system $(x^\alpha)=(x^1,x^2,x^3,x^4)$; spatial indices $i,j,k,\ldots$ will run from $1$ to $3$. The NC connection, $\nabla$, of interest to us is given by its only non-vanishing coefficients
\begin{equation}
\Gamma^i_{44}=-g^i
\label{Gamma}
\end{equation}
which represent the components of the gravitational acceleration, $\bg$, in the chosen co\-ordinate system. 

\goodbreak

The NC gravitational field equations \cite{Kun} are given geometrically in terms of the curvature tensor, $R$, of $\nabla$ by\footnote{\label{convention} Our convention is $R^\delta_{\alpha\beta\gamma}=2\partial_{[\alpha}\Gamma^\delta_{\beta]\gamma}+2\Gamma^\delta_{\kappa[\alpha}\Gamma^\kappa_{\beta]\gamma}$.}
%\footnote{The convention for the sign of the cosmological constant, $\Lambda$, might be here opposite to the usual one.}
\begin{equation}
R_{\alpha\beta}=(4\pi G\varrho+\Lambda)\,\theta_\alpha\theta_\beta
\qquad
\&
\qquad
\gamma^{\kappa\gamma}R^\delta_{\alpha\kappa\beta}
=
\gamma^{\kappa\delta}R^\gamma_{\beta\kappa\alpha}
\label{Newton}
\end{equation}
for all $\alpha,\beta,\gamma,\delta=1,\ldots,4$.
The Ricci tensor, $\Ric$, has components $R_{\alpha\beta}=R^\kappa_{\kappa\alpha\beta}$. In (\ref{Newton}), we denote by $\varrho$ the mass density of the sources and by $\Lambda$ the cosmological constant. These equations imply that $R_{ij}=0$, i.e., space at time $t$ is Ricci-flat, hence (locally) flat. 

In view of (\ref{Gamma}) and (\ref{Newton}), we have
\begin{equation}
\Div\bg=-(4\pi G\varrho+\Lambda)
\qquad
\&
\qquad
\Rot\bg=0
\label{NewtonBis}
\end{equation}
where $\Div$ and $\Rot$ are the standard divergence and curl on $\bR^3$. 

%%%%%%%%%%%%%%%%%%%%%%%%%%%%%%%%%%%%%%%%%%%%%%%%%%%%%%%%%%%%%%%%%%%%%%%%%%
\subsection{Newtonian cosmology: a  review }\label{Ncosmo}
%%%%%%%%%%%%%%%%%%%%%%%%%%%%%%%%%%%%%%%%%%%%%%%%%%%%%%%%%%%%%%%%%%%%%%%%%%

One can trace back the beginning of Newtonian cosmology to work of Kelvin \cite{Kel}. We will also refer to \cite{Mil,MM,ES1,ES2} for the original modern expositions of non-relativistic cosmology. See also the recent contributions \cite{EG1,EG2} using homothetic solutions of the $N$-body problem. 

%-------------------------------------------------------------------------
%\subsubsection{The Cosmological Principle}%\label{CosmologicalPrincipleSection}
%-------------------------------------------------------------------------

\kikezd{\textbf{The Cosmological Principle}}
\medskip

Following \cite{Sou,Duv}, we formulate the cosmological hypothesis of spatial homogeneity and isotropy by demanding that the curvature tensor be rotation-invariant for some freely chosen origin, $\bx=0$, namely that its nonzero components be given by
\begin{equation}
R^i_{j44}=-\partial_jg^i=a\,\delta^i_j.
\label{Rij44=ldeltaij}
\end{equation}
for all $i,j=1,2,3$, and for some function $a$ of time.

\goodbreak

\medskip

\textit{
Condition (\ref{Rij44=ldeltaij}) is strictly equivalent to the vanishing of the pro\-jective Weyl curvature tensor,
\begin{equation*}
W^\delta_{\alpha\beta\gamma}=R^\delta_{\alpha\beta\gamma}-\frac{1}{3}\big(\delta^\delta_\alpha\,R_{\beta\gamma}-\delta^\delta_\beta\,R_{\alpha\gamma}\big)=0
\label{PWtensor}
\end{equation*}
of the NC connection $\nabla$ defined by eqn (\ref{Gamma}). The associated projective geodesic equations~(\ref{hGeodEq1}) are thus invariant under the \textit{local} action of the projective group, $\SL(5,\bR)$, on $4$-dimensional spacetime.\footnote{Generalization to any spacetime dimension $n=d+1>2$ is straightforward \cite{CDGH}. If $n=2$, projective flatness  is guaranteed by the vanishing of the projective Cotton tensor, $\nabla_{\alpha}R_{\beta\gamma}-\nabla_{\alpha}R_{\beta\gamma}=0$; this provides a direct justification of the $\Sl(3,\bR)$ symmetry Lie algebra for the $1$-dimensional damped harmonic oscillator found in \cite{CV}. This local $\SL(n+1,\bR)$ symmetry will eventually be broken by the additional requirement that the Galilei structure, $(\gamma,\theta)$, be conformally preserved as in (\ref{tPhi}).}}

\medskip

Indeed, we see that $W^\delta_{\alpha\beta\gamma}=0$ for the connection~(\ref{Gamma}) iff eqn (\ref{Rij44=ldeltaij}) holds true. Remarkably enough, this implies that the Ricci tensor is actually of the form~(\ref{Newton}) corresponding to the NC field equations. Besides, the fact that $da\wedge\theta=0$, i.e., that $a$ in~(\ref{Rij44=ldeltaij}) is an otherwise arbitrary function of time only is a direct consequence of the Bianchi identities.

Returning to eqn (\ref{Rij44=ldeltaij}), we choose $\bg=0$ at $\bx=0$ to end up with
\begin{equation}
\bg=-a\,\bx.
\label{g}
\end{equation}
Note that (\ref{NewtonBis}) is compatible with uniform expansion velocity, $\bv=d\bx/dt\;(=\dot\bx)$, of the galaxies (\textit{Hubble law}), namely
\begin{equation*}
\bv=H\bx
\label{Hubble}
\end{equation*}
where $H=H(t)$ is the time dependent Hubble parameter. 

\goodbreak

Assuming that $\Lambda=0$ for simplicity, we see that eqns (\ref{NewtonBis}) and~(\ref{g}) readily yield 
\begin{equation}
3a=4\pi G\varrho.
\label{3a}
\end{equation}
As to the principle of geodesics, $\dot\bv=\bg$, applied to freely falling galaxies, it gives then
\begin{equation}
\dot H+H^2+a=0.
\label{dotH}
\end{equation}
At last mass conservation, $\partial_t\varrho+\Div(\varrho\bv)=0$, furthermore implies
\begin{equation}
\dot{a}+3Ha=0.
\label{dota}
\end{equation}
The general solution of this system will be worked out in the next section.

%-------------------------------------------------------------------------
%\subsubsection{The scale factor in terms of cosmological time}
%-------------------------------------------------------------------------

\kikezd{\textbf{The scale factor in terms of cosmological time}}
\medskip

Let us introduce the function $\Theta(t)>0$ on $T$, defined by
\begin{equation}
a(t)=\frac{B}{\Theta(t)^3}
\label{ell}
\end{equation}
where $B=\const>0$ is related to the density of the dust of galaxies. Note that $\Theta$ plays, in Newtonian cosmology, the r\^ole of the scale factor of the universe in the standard model of relativistic cosmology.\footnote{Note that $\Theta(t)$ is interpreted by Souriau \cite{Sou} as the ``co-temperature'' of the cosmological model.}

It is useful and classical to redefine time as follows
\begin{equation}
\tau=\int{\!\!\frac{dt}{\Theta}}
\label{tau}
\end{equation}
which we call ``cosmological time'' (akin to ``conformal time'' of FLRW cosmology).

\goodbreak

Putting $\Theta'=d\Theta/d\tau$, etc., we can transcribe the previous set (\ref{3a})--(\ref{dota}) of coupled equations ruling Newtonian cosmology as
\begin{equation}
H=\frac{\Theta'}{\Theta^2},
\qquad
\Theta'^2-\Theta\Theta''=B\Theta,
\qquad
\frac{2B}{\Theta}-\left(\frac{\Theta'}{\Theta}\right)^2=K.
\label{Friedmann}
\end{equation}
These equations are strictly equivalent to the \textit{Friedmann equations} in the absence of radiation (where~$K$ was interpreted as the spatial constant curvature); here, $K$ is a mere first-integral of the above differential system 
(\ref{dotH}) and (\ref{dota}).
%of Section \ref{CosmologicalPrincipleSection}.
Straightforward integration with the proviso that the scale factor vanishes
%temperature be infinite 
at the big bang, $\tau=0$, yields
\begin{equation}
%\bigbox{
\Theta=\frac{B(1-\cos(k\tau))}{K}
\qquad
\&
\qquad
k=
\left\{
\begin{array}{rcl}
%k=
\sqrt{K}&\hbox{if}&K\geq0\\[6pt]
%k=
i\sqrt{-K}&\hbox{if}&K<0.
\end{array}
\right.
%}
\label{theta}
\end{equation}

%-------------------------------------------------------------------------
%\subsubsection{The scale factor in terms of Newtonian time}
%-------------------------------------------------------------------------

\kikezd{\textbf{The scale factor in terms of Newtonian time}}
\medskip

We note for further use that (\ref{Friedmann}) implies
\begin{equation}
%\medbox{
\Theta^2\ddot\Theta=-B
%}
\label{ddottheta}
\end{equation}
in terms of absolute Newtonian time, $t$, where $\dot\Theta=d\Theta/dt$. This implies that the gravitational potential arising from (\ref{g}) can be chosen as given by
\begin{equation}
%\medbox{
U(\bx,t)=\half\,a(t)\vert\bx\vert^2
%\qquad
%\hbox{where}
%\qquad
%a=\frac{B}{\Theta^3}
%}
\label{acosmo}
\end{equation}
where the function $a$ is as in (\ref{ell}).

%%%%%%%%%%%%%%%%%%%%%%%%%%%%%%%%%%%%%%%%%%%%%%%%%%%%%%%%%%%%%%%%%%%%%%%%%%
%%%%%%%%%%%%%%%%%%%%%%%%%%%%%%%%%%%%%%%%%%%%%%%%%%%%%%%%%%%%%%%%%%%%%%%%%%
\section{Conformal Newton-Cartan symmetries}\label{NC-conformalSection}
%%%%%%%%%%%%%%%%%%%%%%%%%%%%%%%%%%%%%%%%%%%%%%%%%%%%%%%%%%%%%%%%%%%%%%%%%%
%%%%%%%%%%%%%%%%%%%%%%%%%%%%%%%%%%%%%%%%%%%%%%%%%%%%%%%%%%%%%%%%%%%%%%%%%%

In this section we review and substantially extend  several results \cite{Duv} related to non-relativistic ``conformal symmetries'' of classical mechanics in a Newton-Cartan framework. See also \cite{DH-NC,DH}.

%%%%%%%%%%%%%%%%%%%%%%%%%%%%%%%%%%%%%%%%%%%%%%%%%%%%%%%%%%%%%%%%%%%%%%%%%%
\subsection{General definitions}
%%%%%%%%%%%%%%%%%%%%%%%%%%%%%%%%%%%%%%%%%%%%%%%%%%%%%%%%%%%%%%%%%%%%%%%%%%

%-------------------------------------------------------------------------
%\subsubsection{Conformal-Galilei symmetries}
%-------------------------------------------------------------------------

\kikezd{\textbf{Conformal-Galilei symmetries}}
\medskip

Newton-Cartan structures possess ``conformal-projective'' symmetries in the following sense.

%\goodbreak

Just as in the relativistic framework where the group of conformal diffeo\-morphisms (strictly  speaking a pseudogroup because it consists in general of \emph{local} diffeomorphisms) consists of those spacetime transformations preserving the direction of the metric, we will define conformal-Galilei transformations as those (local) diffeomorphisms~$\tPhi$ of $N$ which preserve \textit{independently} the direction of the ``spatial-metric'', $\gamma$, and that of the clock, $\theta$, namely such that
\begin{equation}
\medbox{
\tPhi^*\gamma=f\,\gamma
\qquad
\&
\qquad
\tPhi^*\theta=g\,\theta
}
\label{tPhi}
\end{equation}
for some strictly positive functions $f$ and $g$ of $N$ depending on $\tPhi$. 

These transformations form an \textit{infinite-dimensional} group which we call the \textit{conformal-Galilei group}, $\Conf(N,\gamma,\theta)$.\footnote{It is related to the so-called (centreless) \textit{Schr\"odinger-Virasoro group} \cite{Hen,UR} which is the subgroup $\SV(N,\gamma,\theta)\subset\Conf(N,\gamma,\theta)$ that preserves the direction of $\gamma\otimes\theta$, the special conformal-Galilei structure. The group $\SV(N,\gamma,\theta)$ is thus defined by the additional constraint $\tPhi^*(\gamma\otimes\theta)=\gamma\otimes\theta$ or, in view of (\ref{tPhi}),
\begin{equation}
f\,g=1
\label{fg=1}
\end{equation}
which is the hallmark of Schr\"odinger symmetry (cf. eqn (\ref{z=2})). 
}

\goodbreak

%-------------------------------------------------------------------------
%\subsubsection{Projective symmetries}%\label{ProjSymSection}
%-------------------------------------------------------------------------

\kikezd{\textbf{Projective symmetries}}
\medskip

At this stage, we need to appeal to the symmetries of the equations of free fall according to Cartan and forerunners.

If we denote by $\Gamma^\gamma_{\alpha\beta}$ with $\alpha,\beta,\gamma=1,\ldots,4$ the components of a NC connection, $\nabla$, in 
some
%a chosen 
coordinate system, the equations of free fall read
\begin{equation*}
\frac{d^2x^\gamma}{ds^2}+\Gamma^\gamma_{\alpha\beta}\frac{dx^\alpha}{ds}\frac{dx^\beta}{ds}
=
0
\label{GeodEq}
\end{equation*}
for all $\gamma=1,\ldots,4$, and for some choice of affine parameter, $s$. The choice of an affine  parameter is arbitrary, and we may choose another non-affine parameter $\hs=f(s)$ with $f'(s)\neq0$ to describe the same physics. This reparametrization leaves us with the same geodesics (as embedded one-dimensional submanifolds of $M$) but with another differential equation for them to satisfy, namely
\beq
0
=
\frac{d^2x^\gamma}{d\hs^2}+\Gamma^\gamma_{\alpha\beta}\frac{dx^\alpha}{d\hs}\frac{dx^\beta}{d\hs}+\left[2\Pi_\alpha\frac{dx^\alpha}{d\hs}\right]\frac{dx^\gamma}{d\hs}=
\frac{d^2x^\gamma}{d\hs^2}+\hGamma^\gamma_{\alpha\beta}\frac{dx^\alpha}{d\hs}\frac{dx^\beta}{d\hs}
\label{hGeodEq1}
\eeq
%\gb{unified two displayed formulae}
where 
\begin{equation}
%\medbox{
\hGamma^\gamma_{\alpha\beta}=\Gamma^\gamma_{\alpha\beta}+\delta^\gamma_\alpha\,\Pi_\beta+\delta^\gamma_\beta\,\Pi_\alpha
%}
\label{ProjEquiv}
\end{equation}
with $\Pi$ a $1$-form of $M$; in the above calculation, we find explicitly 
$$
2\Pi_\alpha\frac{dx^\alpha}{d\hs}=\frac{d^2\hs}{ds^2}\left(\frac{ds}{d\hs}\right)^2.
$$

We see from (\ref{hGeodEq1}) that, in terms of the old connection coefficients, $\Gamma^\gamma_{\alpha\beta}$, the new covariant accelera\-tion is proportional to the velocity; this happens to be reminiscent of \textit{friction}, whose systematic study is at the root of the article \cite{CDGH}. The connections~$\nabla$ and~$\hat\nabla$ in (\ref{ProjEquiv}) yielding the same (unparametrised) geodesics are called ``projectively equivalent''~\cite{Mat}. Let us refer to, e.g., \cite{AA} for an account on projective geometry applied to systems of ODE, and also to \cite{Nur,GW} for a discussion about projective symmetries in FLRW cosmology.

\goodbreak

From the physical standpoint, considering unparametrised geodesics as the worldlines of (spinless) particles is the natural way to deal with the Equivalence Principle in the sense of universality of free fall. Therefore the equivalence (\ref{ProjEquiv}) will be considered fundamental, notably for the  definition below.

%-------------------------------------------------------------------------
%\subsubsection{Conformal-NC symmetries}
%-------------------------------------------------------------------------

\kikezd{\textbf{Conformal-NC symmetries}}
\medskip

To comply with the above mentioned Equivalence Principle, and hence restrict ourselves to a finite-dimensional group of spacetime symmetries, we impose that $\tPhi\in\Conf(N,\gamma,\theta)$ be also a projective transformation of NC spacetime, i.e., a transformation that permutes \textit{unparametrised} geodesics, namely such that, according to (\ref{ProjEquiv}), we have
\begin{equation}
\medbox{
\left(\tPhi^*\Gamma\right)_{\alpha\beta}^\gamma=\Gamma_{\alpha\beta}^\gamma
+\delta^\gamma_\alpha\,\Pi_\beta+\delta^\gamma_\beta\,\Pi_\alpha
}
\label{tPhiBis}
\end{equation}
for some $1$-form $\Pi$ depending on $\tPhi$.
One proves easily that necessarily $\Pi=h\,\theta$, where $h$ is some function of $T$. 

\medskip

\textit{
These transformations satisfying (\ref{tPhi}) and (\ref{tPhiBis}) generate what we choose to call the ``conformal-NC group'', and denote by $\Chr(N,\gamma,\theta,\nabla)$; cf. \cite{Duv,BDP0,BDP,Duv2}.\footnote{We find that this group projects to the time axis as a group of projective transformations, i.e., diffeomorphisms of $T$ with vanishing Schwarzian derivative.}
}

\medskip

\goodbreak

This supergroup of the Schr\"odinger group has recently been studied and named the \textit{enlarged Schr\"odinger group} \cite{Gun}. 

The infinitesimal version of the preceding definitions (\ref{tPhi}) and (\ref{tPhiBis}) is the following: we will call \textit{conformal-NC vector field}, $X\in\chr(N,\gamma,\theta,\nabla)$, any vector field $X$ of $N$, solution of the following system \cite{Duv,Duv2}
\begin{equation}
\medbox{
L_X\gamma=\lambda\,\gamma,
\qquad
L_X\theta=\mu\,\theta,
\qquad
L_X\Gamma_{\alpha\beta}^\gamma=
\nu\left(\delta^\gamma_\alpha\,\theta_\beta+\delta^\gamma_\beta\,\theta_\alpha\right)
}
\label{X}
\end{equation}
for some functions $\lambda,\mu,\nu$ of the time axis, $T$, depending on $X$. Let us also refer to \cite{DH-NC,DH}.
We find that, quite generally,
\begin{equation}
\lambda+\mu=\const
\qquad
\&
\qquad
\nu=\half\dot\mu\,.
\label{lambdamunu}
\end{equation}

%%%%%%%%%%%%%%%%%%%%%%%%%%%%%%%%%%%%%%%%%%%%%%%%%%%%%%%%%%%%%%%%%%%%%%%%%%
\subsection{Some examples}\label{Examples}
%%%%%%%%%%%%%%%%%%%%%%%%%%%%%%%%%%%%%%%%%%%%%%%%%%%%%%%%%%%%%%%%%%%%%%%%%%

We will now determine the Lie algebras spanned by the solutions $X=X^i(x)\partial_i+X^4(x)\partial_4$ of~(\ref{X}) in various instances of NC structures. Using standard notation, we recall that
\begin{subequations}
\begin{align}
\label{LXgamma}
L_X\gamma^{\alpha\beta}
&=
-2\gamma^{\kappa(\alpha}\nabla_\kappa{}X^{\beta)}\\[4pt]
\label{LXtheta}
L_X\theta_{\alpha}
&=
\theta_\kappa\nabla_\alpha{}X^\kappa\\[4pt]
\label{LXGamma}
L_X\Gamma_{\alpha\beta}^\gamma
&=
\nabla_\alpha\nabla_\beta{}X^\gamma+R^\gamma_{\kappa\alpha\beta}X^\kappa.
\end{align}
%\label{recall}
\end{subequations}

%-------------------------------------------------------------------------
%\subsubsection{The flat Galilean case}
%-------------------------------------------------------------------------

\kikezd{\textbf{The flat Galilean case}}
\medskip

Let us start with the case of a flat affine connection, $\Gamma_{\alpha\beta}^\gamma=0$, 
%in the above-chosen coordinate system on 
of $\bR^4$. In view of (\ref{LXgamma})--(\ref{LXGamma}), the only non-trivial eqns (\ref{X}) are 
%in this case
%\begin{subequations}
%\begin{align}
%\label{25}
%\partial_iX_j+\partial_jX_i&=-\lambda\,\delta_{ij}\\
%%\label{26}
%%\partial_iX^4&=&0\\
%\label{27}
%\partial_4X^4&=\mu\\
%%\label{28}
%%\partial_i\partial_jX^k&=&0\\
%%\label{29}
%%\partial_i\partial_jX^4&=&0\\
%\label{210}
%\partial_4\partial_jX^k&=\nu\,\delta_j^k
%%\\
%%\label{211}
%%\partial_4\partial_4X^k&=&0\\
%%\label{212}
%%\partial_4\partial_4X^4&=&0
%\end{align}
%\label{galcase}
%\end{subequations}
%%where all those with zero right-hand side have not been displayed.
\beq
\label{25-27-210}
\partial_iX_j+\partial_jX_i=-\lambda\,\delta_{ij},
\qquad
\partial_4X^4=\mu,
\qquad
\partial_4\partial_jX^k=\nu\,\delta_j^k.
\eeq
We find that 
$\nu=\half\dot\mu=-\half\dot\lambda$, so that (\ref{25-27-210}) and
%hence (\ref{210}) and 
%(\ref{211})
$\partial_4\partial_4X^k=0$
imply $\ddot\mu=0$. 
We are thus left with
$
\lambda=-2\alpha{}t-2\chi,
\mu=2\alpha{}t+\eta,
\nu=\alpha
$
where $\alpha,\chi,\eta\in\bR$. 
At last, the general solution of~(\ref{X}) is given in this case by
%\begin{eqnarray}
\begin{subequations}
\begin{align}
\label{Xi}
X^i&=\omega^i_jx^j+\beta^it+\gamma^i+\alpha{}x^i{}t+\chi{}x^i
\\[4pt]
\label{X4}
X^4&=\alpha{}t^2+\eta{}t+\varepsilon
%\end{eqnarray}
\end{align}
\label{XiX4}
\end{subequations}
for all $i=1,2,3$, where $\omega\in\sor(3)$; $\bbeta,\bgamma\in\bR^3$; $\alpha,\chi,\eta,\varepsilon\in\bR$.\footnote{These infinitesimal generators are interpreted as follows: $\omega$ is a rotation, $\bbeta$ a Galilei boost, $\bgamma$ a space translation, $\alpha$ a time inversion, $\chi$ a space dilation, $\eta$ a time dilation, and $\varepsilon$ a time translation. We notice that space and time dilations are \textit{de facto} independent.}

Therefore $\chr(\bR^4,\gamma,\theta,\nabla)$ is generated (with a slight abuse of notation) by the Lie algebra of matrices
\begin{equation}
%\bigbox{
X=\left(
\begin{array}{ccc}
\omega&\bbeta&\bgamma\\
0&\eta-\chi&\varepsilon\\
0&-\alpha&-\chi
\end{array}
\right)\in\chr(\bR^4).
%}
\label{chrR4}
\end{equation}

\medskip
\textit{
We have thus proved that the ``conformal-NC algebra'' is isomorphic to
\begin{equation}
\medbox{
\chr(\bR^4)\cong(\sor(3)\times\gl(2,\bR))\ltimes(\bR^3\times\bR^3).
}
\label{chr4}
\end{equation}
}
\medskip 

Generalization of (\ref{chr4}) to $n$-dimensional NC spacetime is immediate. As to the ``Schr\"odinger Lie algebra''
\begin{equation}
\sch(\bR^4)\cong(\sor(3)\times\Sl(2,\bR))\ltimes(\bR^3\times\bR^3)
\label{sch4}
\end{equation}
it corresponds, in (\ref{chrR4}), to the choice
\begin{equation}
\eta=2\chi.
\label{eta=2chi}
\end{equation}

Let us emphasize that preservation of the equations of free fall (which in essence do not involve the mass of the test particle) leaves the dilation factors $\chi$ and $\eta$ totally independent. But eqn (\ref{eta=2chi}) is in fact designed to guarantee mass conservation in the Schr\"odinger equation \cite{Duv,Duv2,DGH}. Thus, the infinitesimal Schr\"odinger dilation is of the form
\begin{equation*}
X=x^i\partial_i+2t\,\partial_t
\label{schDilation}
\end{equation*}
 and features the \textit{dynamical exponent} 
\begin{equation}
z=2.
\label{z=2}
\end{equation}
The \textit{Schr\"odinger Lie algebra} (\ref{sch4}) is a finite-dimensional Lie subalgebra of the Schr\"odinger-Virasoro Lie algebra, defined by $\lambda+\mu=0$ (cf. (\ref{lambdamunu})), expressing infinitesimally the constraint (\ref{fg=1}).

%-------------------------------------------------------------------------
%\subsubsection{Newtonian cosmology}%\label{NewtonianCosmologySymmetriesSection}
%-------------------------------------------------------------------------

\kikezd{\textbf{Newtonian cosmology}}

\medskip

Taking into account eqn (\ref{Gamma}) defining our NC spacetime structure, we see that the general solution of (\ref{X}) is again given by 
%the system 
(\ref{25-27-210})
%(\ref{25})--(\ref{210});
%and (\ref{212}), i.e., $\partial_4\partial_4X^4=0$; 
but with 
%(\ref{211}) 
$\partial_4\partial_4X^k=0$ replaced by
\begin{equation}
\partial_4\partial_4X^k-X^4\partial_4g^k-2g^k\partial_4X^4+g^j\partial_jX^k-X^j\partial_jg^k=0
\label{LXGammak44}
\end{equation}
where the expression (\ref{LXGamma}) of the Lie derivative of a connection has been used.

\goodbreak

From %(\ref{26}) the fact that
$\partial_iX^4=0$ 
and using %eqns (\ref{27}), (\ref{g}) and 
eqns (\ref{LXGammak44}) and (\ref{g}) we get 
$
\dddot{X}^4+2\dot{a}\,X^4+4a\,\dot{X}^4=0.
$
Expressing this in term of the new parameter $\tau$, cf. (\ref{tau}), and with the help of (\ref{Friedmann}), we obtain 
$$
\Theta(X^4)'''+(9B-2K\Theta)(X^4)'-3\Theta'(X^4)''-6B(\Theta'/\Theta)X^4=0.
$$ 
More effort is needed to deduce that
$(X^4/\Theta)'''+K(X^4/\Theta)'=0$, hence that
\begin{equation}
\medbox{
X^4=\Theta\left[2\alpha\,\frac{(1-\cos(k\tau))}{K}+\eta\,\frac{\sin(k\tau)}{k}+\varepsilon\right]
}
\label{X4cosmo}
\end{equation}
where $\Theta$ is as in (\ref{theta}) and $\alpha,\eta,\varepsilon\in\bR$. 

\goodbreak

We note that the expression (\ref{X4cosmo}) admits an analytical extension to the case $K<0$, and is well-defined in the limit $K\to0$, see~(\ref{theta}).

The equation  $\partial_i\partial_jX^k=0$
then tells us that $x^i=\Omega^i_jx^j+\Gamma^i$, for some $\Omega$ and $\bGamma$ depending on $t$ only. Now (\ref{25-27-210}) implies that 
$$
\Omega^i_j=\omega^i_j+(\half\mu+a)\delta^i_j
$$
 with $\omega\in\sor(3)$ and $a\in\bR$. Then,  eqn (\ref{LXGammak44}) gives $\partial_4\partial_4X^i=\half\ddot\mu\,x^i-a\,\Gamma^i=\half\ddot\mu\,x^i+\ddot\Gamma^i$ from the preceding expression of $X^i$. We thus find $\ddot\Gamma^i+a\,\Gamma^i=0$, hence $(\Gamma^i)'''+K(\Gamma^i)'=0$. Bearing in mind that $\Theta=\Theta'=0$ at $\tau=0$, we get 
$$
\Gamma^i=\beta^i\frac{(1-\cos(k\tau))}{K}+\gamma^i\frac{\sin(k\tau)}{2k}
$$ 
for some vectorial constants of integration $\bbeta$ and $\bgamma$. The next step consists in showing, via~(\ref{X4cosmo}), that eqn (\ref{25-27-210}) yields the clock dilation factor in~(\ref{X}), namely
%\begin{equation}
$$
\mu=\eta\,(1+2\cos(k\tau))+4\alpha\,\frac{\sin(k\tau)}{k}+\varepsilon\,\frac{k}{\tan(k\tau/2)}.
$$
%\label{mu}
%\end{equation}
Redefining several constants of integration allows us to conclude that
\begin{equation}
\medbox{
\begin{array}{rcl}
X^i
&=&
\displaystyle
\omega^i_jx^j+\left[\eta\,\cos(k\tau)+2\alpha\,\frac{\sin(k\tau)}{k}
+\varepsilon\,\frac{k/2}{\tan(k\tau/2)}+\chi\right]x^i\\[12pt]
&&
\displaystyle
+\beta^i\,\frac{(1-\cos(k\tau))}{K}
+\gamma^i\,\frac{\sin(k\tau)}{2k}
\end{array}
}
\label{Xicosmo}
\end{equation}
for all $i=1,2,3$, where $\omega\in\sor(3)$; $\bbeta,\bgamma\in\bR^3$; $\alpha,\chi,\eta,\varepsilon\in\bR$. Again, this expression can be extended to the case $K\leq0$; cf. (\ref{theta}).

\medskip

\textit{
The vector fields 
$$
X=X^i(x)\partial_i+X^4(t)\partial_4\in\chr(N,\gamma,\theta,\nabla)
$$ 
defined by (\ref{X4cosmo}) and (\ref{Xicosmo}) span the $13$-dimensional conformal-NC Lie algebra of the cosmological models which has been shown \cite{Duv} to be isomorphic to the Lie algebra (\ref{chr4}) of the matrices (\ref{chrR4}).
}

\goodbreak

%-------------------------------------------------------------------------
%\subsubsection{Vacuum with cosmological constant (Newton-Hooke symmetry)}
%\label{Cosmo-NH-SymmetrySection}
%-------------------------------------------------------------------------

\kikezd{\textbf{Vacuum with cosmological constant (Newton-Hooke symmetry)}}

\medskip

We start, in this section, with the general spherical-symmetric vacuum solution of the field equations~(\ref{NewtonBis}), i.e., $\varrho=0$, and $\Lambda\neq0$, namely
\begin{equation*}
\bg=-\left(\frac{K}{r^3}+\frac{\Lambda}{3}\right)\bx
\label{gKLambda}
\end{equation*}
with $K=G\cM=\const>0$ (with $\cM$ the mass of the pointlike source), and $r=\vert\bx\vert>0$. Here $N=\bR^3\setminus\{0\}\times\bR$, while the Galilei tensors $\gamma$ and $\theta$ are as before. 

\goodbreak

Putting 
$
k=\sqrt{\Lambda/3}
$
we find, via (\ref{X}), that
\begin{equation}
\mu=\alpha\,\frac{\sin(2k t)}{k}+\eta\,\cos(2k t)
\label{muKLambda}
\end{equation}
with $\alpha,\eta\in\bR$. This expression is still well-defined in the case $\Lambda\leq0$. 

Using (\ref{lambdamunu}), we will write $\lambda=-\mu+\eta-2\chi$ with $\chi\in\bR$. We find moreover that $X^i=\omega^i_jx^j+\half(f+2\chi-\eta)x^i+\Gamma^i$, where $\omega\in\sor(3)$ and $\ddot\Gamma^i+(\Lambda/3)\Gamma^i=0$.

\medskip

$\bullet$ If $K=0$ \& $\Lambda\neq0$ (no Newtonian gravitational source, $\cM=0$, and non-vanishing cosmo\-logical constant), we easily find 
$$
\Gamma^i=\beta^i\sin(kt)/k+\gamma^i\cos(kt)
$$
for some vectorial constants $\bbeta$ and~$\bgamma$. From the preceding expression of $X^i$ we finally obtain
\begin{equation}
\medbox{
\begin{array}{rcl}
X^i
&=&
\displaystyle
\omega^i_jx^j+\left[\half\eta\,(\cos(2k t)-1)+\alpha\,\frac{\sin(2k t)}{2k}
+\chi\right]x^i\\[12pt]
&&
\displaystyle
+\beta^i\,\frac{\sin(k t)}{k}
+\gamma^i\,\cos(k t)
\end{array}
}
\label{XiLambda}
\end{equation}
and, with help of (\ref{25-27-210}), we end up with
\begin{equation}
\medbox{
X^4=\alpha\,\frac{(1-\cos(2k t))}{2k^2}+\eta\,\frac{\sin(2k t)}{2k}+\varepsilon
}
\label{X4Lambda}
\end{equation}
where $\omega\in\sor(3)$; $\bbeta,\bgamma\in\bR^3$; $\alpha,\chi,\eta,\varepsilon\in\bR$.

\medskip

\textit{
These vector fields generate $\chr(N,\gamma,\theta,\nabla)$, i.e., the $13$-dimensional conformal-NC extension of the $10$-dimensional \textit{Newton-Hooke} Lie algebra \cite{GP}.}\footnote{Note that we recover (\ref{Xi}) and (\ref{X4}), i.e., the Lie algebra $\chr(\bR^4)$, in the limit $\Lambda\to0$.}

\medskip
\goodbreak

To recover the standard notation of the literature, let us introduce the following basis of our Lie algebra, viz.,
\begin{eqnarray*}
J_i&=&\epsilon_{ij}^{\ \ \ell}x^i\partial_\ell\\%[4pt]
K_i&=&\frac{\sin(k t)}{k}\,\partial_i\\%[4pt]
P_i&=&\cos(k t)\,\partial_i\\%[4pt]
H&=&\partial_t\\%[4pt]
I&=&\frac{\sin(2k t)}{2k}\,x^i\partial_i+\frac{(1-\cos(2k t))}{2k^2}\,\partial_t\\%[4pt]
T&=&\half\,(\cos(2k t)-1)x^i\partial_i+\frac{\sin(2k t)}{2k}\,\partial_t\\%[4pt]
S&=&x^\ell\partial_\ell
\end{eqnarray*}
where $i=1,2,3$. Those correspond respectively to infinitesimal rotations, boosts, space-translations, time-translation, inversion, and \textit{independent} time-dilation, and space-dilation.

The non-trivial Lie brackets are as follows:
\begin{equation*}
\begin{array}{llll}
&[J_i,J_j]=-\epsilon_{ij}^{\ \ \ell}J_\ell,
&[J_i,K_j]=-\epsilon_{ij}^{\ \ \ell}K_\ell,
&[J_i,P_j]=-\epsilon_{ij}^{\ \ \ell}P_\ell,\\[4pt]
&[K_i,H]=-P_i,
&[K_i,T]=-K_i,
&[K_i,S]=K_i,\\[8pt]
&[P_i,H]=-k^2K_i,
&[P_i,I]=K_i,
&[P_i,S]=P_i,\\[6pt]
&[H,I]=2T+S,
&[H,T]=-2k^2 I+H,
&[I,T]=-I.
\end{array}
\label{chrNH}
\end{equation*}

\medskip

$\bullet$ If $\Lambda=0$ \& $K>0$ (in the presence of a gravitational source located at $r=0$, with a vanishing cosmological constant), eqn (\ref{muKLambda}) gives $\mu=2\alpha t+\eta$. Straightforward calculation then yields 
$$
2\Gamma^i+\left[\mu-3(2\chi-\eta)-6\frac{\Gamma_jx^j}{r^2}\right]x^i=0;
$$
 but we have seen that $\ddot\Gamma^i=0$ since $\Lambda=0$, hence $\Gamma^i=\beta^i t+\gamma^i$. Compatibility of these expressions implies that $\Gamma^i=0$. We moreover find, using $\mu-3(2\chi-\eta)=0$, that $\alpha=0$ and $3\chi=2\eta$. 
This leaves us with
\begin{equation}
\medbox{
X^i
=
\omega^i_jx^j+\chi\,x^i
\qquad
\&
\qquad
X^4=\frac{3}{2}\chi\,t+\varepsilon
}
\label{XiK}
\end{equation}
where $\omega\in\sor(3)$; $\chi,\varepsilon\in\bR$. 

\goodbreak

\medskip

\textit{
The Lie algebra $\chr(N,\gamma,\theta,\nabla)$ defined by (\ref{XiK}), 
which may be called the ``Virial Lie algebra'', is $5$-dimensional and contains the ``Keplerian dilations'', generated by 
\begin{equation*}
X=x^i\partial_i+\frac{3}{2}\,t\,\partial_t
\end{equation*}
featuring a dynamical exponent $z=\frac{3}{2}$.} 

\medskip
 
These spacetime dilations are precisely the transformations behind Kepler's third law. In the symplectic framework, they generate \textit{canonical similitudes} \cite{SSD} ; as such, they do not generate any Noetherian conserved quantity.

%%%%%%%%%%%%%%%%%%%%%%%%%%%%%%%%%%%%%%%%%%%%%%%%%%%%%%%%%%%%%%%%%%%%%%%%%%
%%%%%%%%%%%%%%%%%%%%%%%%%%%%%%%%%%%%%%%%%%%%%%%%%%%%%%%%%%%%%%%%%%%%%%%%%%
\section{Conformal-Bargmann symmetries}\label{Conf-BargSymmetrySection}
%%%%%%%%%%%%%%%%%%%%%%%%%%%%%%%%%%%%%%%%%%%%%%%%%%%%%%%%%%%%%%%%%%%%%%%%%%
%%%%%%%%%%%%%%%%%%%%%%%%%%%%%%%%%%%%%%%%%%%%%%%%%%%%%%%%%%%%%%%%%%%%%%%%%%

We rephrase now the above results, in a somewhat \textit{simpler} fashion, on $(4+1)$-dimensional Bargmann extended spacetime over a $4$-dimensional Newton-Cartan spacetime. This strategy has been called the \textit{Eisenhart lift} of NC structures in, e.g., \cite{Car,BM,Car2}.

%%%%%%%%%%%%%%%%%%%%%%%%%%%%%%%%%%%%%%%%%%%%%%%%%%%%%%%%%%%%%%%%%%%%%%%%%%
\subsection{General definitions}
%%%%%%%%%%%%%%%%%%%%%%%%%%%%%%%%%%%%%%%%%%%%%%%%%%%%%%%%%%%%%%%%%%%%%%%%%%

Recall that a general Bargmann structure is a triple $(M,\rg,\xi)$ where $M$ is a $m$-dimensional smooth manifold endowed with a Lorentz metric, $\rg$, together with  a distinguished nowhere zero vector field, $\xi$, which is null and parallel \cite{DBKP}. These structures fall into Brinkmann's classification \cite{Bri}. It has also been shown that a Bargmann structure actually projects onto a NC structure $(N,\gamma,\theta,\nabla)$ where $N=M/(\bR\xi)$ is inter\-preted as $n$-dimensional spacetime, where $n=m-1=d+1$. 

We will use this fact to advantage\-ously lift \textit{\`a la} Eisenhart \cite{Eis} the above studied NC sym\-metries to the corresponding Bargmann structure. This will help us desingularise the Galilei ``metric'' structures on  $N$, while  recovering the familiar Lorentzian geometry on an extended spacetime $M$. This procedure has been systematically exploited in, e.g., \cite{DGH,BDP,Car,BM,DL}.

\medskip
\textit{
The conformal-NC group admits a $(\bR,+)$-extension as the ``conformal-Bargmann group'' defined as the ``group'' $\Chr(M,\rg,\xi)$ of all (local) diffeomorphisms, $\Phi$, of~$M$ such that
\begin{equation}
\medbox{
\Phi^*\rg=\lambda\,\rg
\qquad
\&
\qquad
\Phi^*\xi=\nu\,\xi
}
\label{ChrBargmann}
\end{equation}
for some smooth, positive, real functions $\lambda,\nu$ of $M$ \cite{BDP,DL}.
}

\medskip

One easily checks that necessarily
\begin{equation*}
d\lambda\wedge\theta=0
\qquad
\&
\qquad
d\nu=0
\label{lambdanu}
\end{equation*}
which means that $\lambda$ is actually a function of time, $T=M/\xi^\perp$, and $\nu$ a constant. 

\goodbreak

Since $\Phi$ permutes the fibres of the principal $(\bR,+)$-bundle $M\to{}N$, it descend to NC spacetime, $N$, as a conformal-NC transformation, $\tPhi$, as in (\ref{tPhi}) and (\ref{tPhiBis}). 
We thus have the exact sequence of group homomorphisms
$$
1\to\bR\to\Chr(M,\rg,\xi)\to\Chr(N,\gamma,\theta,\nabla)\to1.
$$
Let us emphasize that $\Chr(M,\rg,\xi)$ is a non-central extension of $\Chr(N,\gamma,\theta,\nabla)$ as long as $\nu\neq1$ in (\ref{ChrBargmann}).\footnote{\label{nu=1}The $(\bR,+)$ subgroup (of ``vertical translations'') is central in the \textit{Schr\"odinger} subgroup, $\Sch(M,\rg,\xi)$, defined by $\nu=1$ in (\ref{ChrBargmann}).}

%%%%%%%%%%%%%%%%%%%%%%%%%%%%%%%%%%%%%%%%%%%%%%%%%%%%%%%%%%%%%%%%%%%%%%%%%%
%\subsection{Conformal-Bargmann flatness}
%%%%%%%%%%%%%%%%%%%%%%%%%%%%%%%%%%%%%%%%%%%%%%%%%%%%%%%%%%%%%%%%%%%%%%%%%%

Given a Bargmann structure, if $C=(C_{\lambda\mu\nu\rho})$ is the conformal Weyl tensor of $(M,\rg)$, using the extra vector field, $\xi$, we can form the tensor $P$ defined by
\begin{equation}
P_{\lambda\mu\nu}=C_{\lambda\mu\nu\rho}\,\xi^\rho.
\label{P}
\end{equation}
This tensor is clearly invariant under the rescalings (\ref{ChrBargmann}) of the metric, $\rg$, and the ``wave-vector'' $\xi$.
Let us mention that the vanishing (\ref{P=zero}) of the tensor (\ref{P}) will in fact constitute the definition (\ref{CBW}) of the conformal-Bargmann Weyl tensor \cite{Duv2,PBD}. 

\textit{Conformal-Bargmann flatness} means that $(M,\rg,\xi)$ can be locally conformally mapped to $(\bR^{d+2},\rg_0,\xi_0)$ where
\begin{equation}
\rg_0=\delta_{ij}\,dx^i\otimes{}dx^j+dt\otimes{}ds+ds\otimes{}dt.
\qquad
\&
\qquad
\xi_0=\frac{\partial}{\partial{}s}
\label{g0xi0}
\end{equation}
with the usual notation, i.e., there exists a local diffeomorphism $\Psi:M\to\bR^{d+2}$ such that
\begin{equation}
\medbox{
\rg=\Theta^2\,\Psi^*\rg_0
\qquad
\&
\qquad
\xi=\Psi^*\xi_0
}
\label{gg0xixi0}
\end{equation}
for some function $\Theta>0$ necessarily defined on $T$. 

Under these circumstances, one is easily convinced that the conformal-Bargmann groups of these Bargmann structures are isomorphic, namely
\begin{equation*}
\medbox{
\Chr(M,\rg,\xi)\cong\Chr(\bR^{d+2},\rg_0,\xi_0)
}
\label{ChrChr0}
\end{equation*}
and that the isomorphism $\Chr(\bR^{d+2},\rg_0,\xi_0)\to\Chr(M,\rg,\xi)$ is given by 
\begin{equation}
\Phi_0\mapsto\Phi=\Psi^{-1}\circ\Phi_0\circ\Psi.
\label{Phi}
\end{equation}

\goodbreak

Let us recall that the \textit{conformal-Bargmann group} (alias the ``chronoprojective group'') acts on flat Bargmann space (\ref{g0xi0}) as follows~\cite{DL}. The action $\Phi:(\bx,t,s)\mapsto(\tbx,\wtt,\ts)$ of $\Phi\in\Chr(\bR^{d+2},\rg_0,\xi_0)$ reads
\begin{eqnarray}
\label{tbx}
\tbx&=&\frac{A\bx+\bb t+\bc}{f t+g}\\[4pt]
\label{wtt}
\wtt&=&\frac{d t+e}{f t+g}\\[4pt]
\label{ts}
\ts&=&\frac{1}{\nu}\left[
s+\frac{f}{2}\frac{\vert{}A\bx+\bb t+\bc\vert^2}{f t+g}
-\la\bb,A\bx\ra-\half\vert\bb\vert^2t+h
\right]
\end{eqnarray}
where $A\in\rO(d)$, $(\bb,\bc)\in\bR^d\times\bR^d$ (boosts \& space-translations), also
\begin{equation*}
%D=
\left(
\begin{array}{cc}
d&e\\f&g
\end{array}
\right)\in\GL(2,\bR)
\label{GL2}
\end{equation*}
is a projective transformation of the time-axis, and $h\in\bR$ (extension parameter). We note that that
$\lambda=(f\,t+g)^{-2}$ and 
$\nu=dg-ef$.\footnote{We find that $\dim\left(\Chr(\bR^{d+2},\rg_0,\xi_0\right)=\half(d^2+3d+10)$, which is $14$ if $d=3$.}

According to an old suggestion of Souriau's, the subgroup defined by the constraints $f=0,d=g=\nu=1$ is now called the  \textit{Bargmann group} after \cite{Bar}; it is the centrally extended Galilei group geometrically defined as the group of strict auto\-morphisms of a Bargmann structure $(M,\rg,\xi)$.

\goodbreak

Let us now emphasize that the following statements are equivalent \cite{PBD} (see also \cite{BDP0} and \cite{Duv} for an approach based on NC theory).
\begin{enumerate}
\item
The Bargmann structure admits a second-order \textit{normal} Cartan connection for the conformal-Bargmann group.
\item
The Bargmann structure is of Petrov type N (see (\ref{Cxi=0})), i.e., 
\begin{equation}
P_{\lambda\mu\nu}=0.
\label{P=zero}
\end{equation}
\item
There holds
\begin{equation}
R_{\mu\nu}=F\,\xi_\mu\xi_\nu
\label{Ric}
\end{equation}
with $F$ some arbitrary function of spacetime, $N$.\footnote{Note that $F$ is a priori a function of $M$; now the Einstein tensor being divergence free, we get 
$\xi(F)=0$, and $F$ is hence (the pull-back of) a function of spacetime, $N$.}
\end{enumerate}

Putting $F=4\pi{}G\,\varrho+\Lambda$ enables us to interpret (\ref{Ric}) as the \textit{Newton-Cartan gravitational field equations}, see (\ref{Newton}), which show up in this remarkable guise in the Bargmann framework.

It has been proved in \cite{Duv2,PBD}, using the fact that $\xi$ is null and parallel (implying  $R_{\lambda\mu\nu\rho}\,\xi^\rho=0$ and $R_{\mu\nu}\,\xi^\nu=0$), that the above condition (\ref{P=zero}) yields 
\begin{equation}
\medbox{
\begin{array}{c}
C_{\lambda\mu\nu\rho}=R_{\lambda\mu\nu\rho}
-
\displaystyle
\frac{1}{n-2}\big(
\rg_{\lambda\nu}\,R_{\mu\rho}
-
\rg_{\lambda\rho}\,R_{\mu\nu}
-
\rg_{\mu\nu}\,R_{\lambda\rho}
+
\rg_{\mu\rho}\,R_{\lambda\nu}
\big)
\\[6pt]
\&
\\[4pt]
R_{\mu\nu}\,\xi_\rho=R_{\mu\rho}\,\xi_\nu
\end{array}
}
\label{CBW}
\end{equation}
where $n=d+2$ is the dimension of $M$. This tensor $C$ (where the Ricci tensor hence satisfies (\ref{Ric})) is called the \textit{Bargmann-conformal Weyl tensor}.

\medskip
\textit{
The Bargmann structure is therefore (locally) conformally-Bargmann flat iff}
\begin{equation}
C_{\lambda\mu\nu\rho}=0.
\label{C=0}
\end{equation}
%\medskip

%%%%%%%%%%%%%%%%%%%%%%%%%%%%%%%%%%%%%%%%%%%%%%%%%%%%%%%%%%%%%%%%%%%%%%%%%%
\subsection{Bargmann-conformally flat examples}
%%%%%%%%%%%%%%%%%%%%%%%%%%%%%%%%%%%%%%%%%%%%%%%%%%%%%%%%%%%%%%%%%%%%%%%%%%

Let us consider the case $d=3$ already dealt with in Section \ref{NC-cosmologySection}.

%-------------------------------------------------------------------------
%\subsubsection{The time-dependent harmonic oscillator}
%-------------------------------------------------------------------------

\kikezd{\textbf{The time-dependent harmonic oscillator}}
\medskip

Start with the Bargmann structure $(M,\rg_U,\xi)$, where $M\subset\bR^5$,  and
\begin{equation}
%\medbox{
\rg_U = \rg_0-2Udt\otimes{}dt
\qquad
\&
\qquad
\xi=\frac{\partial}{\partial{}s}
%}
\label{gU}
\end{equation}
with $U$ the gravitational potential. One proves \cite{Duv2} that eqn (\ref{C=0}) holds whenever
$$
U(\bx,t)=\half{}a(t)\vert\bx\vert^2+\bb(t)\cdot\bx+c(t)
$$
for some arbitrary functions $a,\bb,c$ of the time-axis,~$T$.  More general situations were studied in \cite{DHP}.

\goodbreak

%-------------------------------------------------------------------------
%\subsubsection{Worked examples of chronoprojective flatness}
%-------------------------------------------------------------------------

Let us stick henceforth to the interesting case
\begin{equation}
U(\bx,t)=\half{}a(t)\vert\bx\vert^2
\label{U}
\end{equation}
and posit, a priori, $\Psi(\bx,t,s)=(\hbx,\wht,\hs)$ where
%\begin{eqnarray}
%\label{hbx}
%\hbx&=&\frac{\bx}{\theta}\\[6pt]
%\label{wht}
%\wht&=&\int{\!\frac{dt}{\theta^2}}\\
%\label{hs}
%\hs&=&s+\half\frac{\dot\theta}{\theta}\vert\bx\vert^2
%\end{eqnarray}
\begin{equation}
\hbx=\frac{\bx}{\Theta},
%\qquad
\qquad
\wht=\int{\!\frac{dt}{\Theta^2}},
%\qquad
\qquad
\hs=s+\half\frac{\dot\Theta}{\Theta}\vert\bx\vert^2
\label{xtohx}
\end{equation}
with $\Theta$ an arbitrary function of time $t$. 

\goodbreak

A short calculation leads us to eqn (\ref{gg0xixi0}) for $\rg=\rg_U$ and $\xi=\partial/\partial{}s$ (in the present case) where the potential (\ref{U}) is such that \cite{BDP}
\begin{equation}
a(t)=-\frac{\ddot\Theta(t)}{\Theta(t)}.
\label{a}
\end{equation}
This implies that $(M,\rg_U,\xi)$ is conformally-Bargmann flat. Thus $\Phi\in\Chr(M,\rg_U,\xi)$ is given by (\ref{Phi}) where 
\begin{equation}
\Phi_0:(\hbx,\wht,\hs)\mapsto(\tbx,\wtt,\ts)
\label{hat-tilde}
\end{equation}
is as in (\ref{tbx})--(\ref{ts}). 

Let us now illustrate these findings with the help of several special example, namely the cosmological Newtonian models, and the Newton-Hooke spacetime models.

%-------------------------------------------------------------------------
%\subsubsection{Newtonian cosmology}%\label{CosmoSection}
%-------------------------------------------------------------------------

\kikezd{\textbf{Newtonian cosmology}}
\medskip

We have seen that eqns (\ref{Friedmann}) readily imply  (\ref{ddottheta}). In view of (\ref{a}), this implies that
the gravitational potential of Newtonian cosmology in the metric (\ref{gU}) is given by (\ref{U}) with
\begin{equation*}
a(t)=\frac{B}{\Theta^3}
\label{acosmobis}
\end{equation*}
in accordance with (\ref{ell}). We have just confirmed the \textit{conformal-Bargmann symmetry} of the Newtonian cosmological model --- without cosmological constant.

We now exhibit the (inverse of the) diffeomorphism (\ref{xtohx}) that brings (locally) conformal\-ly the cosmological Bargmann model to flat Bargmann space (\ref{gg0xixi0}). In view of (\ref{theta}) and (\ref{tau}), we find 
$$
t=\int{\!\Theta\,d\tau}=B\tau-\frac{B}{k^3}\sin(k\tau)
$$
 up to a constant. And (\ref{xtohx}) yields 
$$
\wht=\int{\!\frac{dt}{\Theta^2}}=\int{\!\frac{d\tau}{\Theta}}
=
-\frac{k}{B}\cot\left(\frac{k\tau}{2}\right)
$$
modulo a constant; see also \cite{HH}. 
This shows that
\begin{equation}
\tau=-\frac{2}{k}\arccot\left(\frac{B\,\wht}{k}\,\right)
\label{taucosmo}
\end{equation}
which helps us express the scale factor
%co\-temperature 
$\Theta=(B/k^2)(1-\cos(k\tau))$ given by (\ref{theta}) as the function
\begin{equation}
\Theta=\frac{2B}{2B^2\,\wht^2+k^2}
\label{thetacosmo}
\end{equation}
of ``time'' $\wht$. 
At last, the sought diffeomorphism $\Psi^{-1}$ (see (\ref{xtohx})) is given by
\begin{equation*}
%\medbox{
\bx=\Theta\,\hbx,
\qquad
%t=\frac{B}{k^2}\left[
%-2\arccot\left(\frac{B\,\wht}{k}\,\right)+\sin\left(2\arccot\left(\frac{B\,\wht}{k}\,\right)\right)\right],
t=B\tau-\frac{B}{k^3}\sin(k\tau),
\qquad
s=\hs+\frac{B\wht}{2}\,\Theta\,\vert\hbx\vert^2
%}
\label{Psicosmo}
\end{equation*}
where $\tau$ is as in (\ref{taucosmo}), and $\Theta$ as in (\ref{thetacosmo}).

\goodbreak

The above action (\ref{tbx})--(\ref{ts}) of the conformal-Bargmann group on flat space $(\bR^{4,1},\xi_0)$ co\-ordinatized by $(\hbx,\wht,\hs)$ is thus transported to $(M,\rg_U,\xi)$ as an action of $\Chr(M,\rg_U,\xi)$. 

\medskip

\textit{
This proves that the conformal symmetry group of Newtonian cosmology lifted to Bargmann's spacetime is (locally) isomorphic to the chronoprojective group $\Chr(\bR^5,\rg_0,\xi_0)$.
}

\medskip

Interestingly, using again (\ref{ddottheta}), we discover that the \textit{Schwarzian derivative} of the diffeomorphism $\phi:t\mapsto\wht$ given by (\ref{xtohx}) is\footnote{The expression (\ref{S}) is no surprise and quite general; it results from the transformation of the Ricci tensor under conformal Bargmann rescalings \cite{DL}.}
\begin{eqnarray}
S(\phi)
&=&\frac{\phi'''}{\phi'}-\frac{3}{2}\left(\frac{\phi''}{\phi'}\right)^2 %\\%[5pt]
=\frac{2B}{\Theta^3}
\label{S}
= 
\frac{8\pi{}G\varrho}{3}
\end{eqnarray}
where $\varrho$ is the time dependent mass density given by the Poisson equation, $\Delta{}U=4\pi{}G\varrho$;
see (\ref{Newton}) with $\Lambda=0$. Now, strictly speaking, the Schwarzian derivative is the quadratic differential $\cS(\phi)=S(\phi)\,dt^2$; the inhomogeneous NC field equations in (\ref{Newton}) can therefore be read off as 
\begin{equation}
\Ric=\frac{3}{2}\cS(\phi).
\label{Ric=S}
\end{equation}

\goodbreak

%-------------------------------------------------------------------------
%\subsubsection{The Hooke solution}
%-------------------------------------------------------------------------

\kikezd{\textbf{The Hooke solution}}
\medskip

This case deals (see (\ref{NewtonBis})) with the gravitational acceleration
$$
\bg=-\frac{\Lambda}{3}\vert\bx\vert^2
$$
hence with the potential (\ref{U}) where
\begin{equation*}
a(t)=k^2
\qquad
\&
\qquad
k=\sqrt{\frac{\Lambda}{3}}\,.
\label{aHooke}
\end{equation*}

\goodbreak

Using (\ref{a}), we find $\Theta=A\cos(k t -\phi)$ for some integration constants $A>0$ and~$\phi$, and (\ref{xtohx}) yields $\wht=\tan(k t - \phi)/(k A^2)$. Now we want $\wht\to{}t$ if $k\to0$, whence the choice
\begin{equation*}
\Theta=\cos(k t)
\label{thetaHooke}
\end{equation*}
which leads to
\begin{equation}
\hbx=\frac{\bx}{\cos(k t)},
\qquad
\wht=\frac{\tan(k t)}{k},
\qquad
\hs=s-\half\,k\tan(k t)\vert\bx\vert^2.
\label{xtohxHooke}
\end{equation}
\textit{
Again $(M,\rg_U,\xi)$ is conformally-Bargmann flat, and the action of $\Chr(M,\rg_U,\xi)$ is as in~(\ref{hat-tilde}) together with (\ref{xtohxHooke}).} 

\medskip

We note also that the formula (\ref{xtohxHooke}) is consistent with the one obtained by turning off the friction term in the ``Arnold'' expression \cite{ACGLR} in eqn (III.76) of \cite{CDGH}.

The infinitesimal version of the corresponding projected action of $\Chr(M,\rg_U,\xi)$ on NC spacetime reproduces eqns (\ref{XiLambda}) and (\ref{X4Lambda}). See also \cite{Nie2}.

At last, the Schwarzian derivative of the diffeomorphism $\phi:t\mapsto\wht$ of the time axis given in (\ref{xtohxHooke}) for the Hooke solution, is
%(using (\ref{aHooke})) that it takes the form
\begin{equation}
S(\phi)=\frac{2}{3}\Lambda
\label{darkEnergy}
\end{equation}
as anticipated in (\ref{S}) and Newton's field equations (\ref{Newton}); compare also  (\ref{Ric=S}). Thus, eqn~(\ref{darkEnergy}) reveals the striking relationship between the Schwarzian derivative and \textit{dark energy} advertised in~\cite{Gib}. 

%%%%%%%%%%%%%%%%%%%%%%%%%%%%%%%%%%%%%%%%%%%%%%%%%%%%%%%%%%%%%%%%%%%%%%%%%%
%%%%%%%%%%%%%%%%%%%%%%%%%%%%%%%%%%%%%%%%%%%%%%%%%%%%%%%%%%%%%%%%%%%%%%%%%%
\section{Conclusion}\label{ConclusionionSection}
%%%%%%%%%%%%%%%%%%%%%%%%%%%%%%%%%%%%%%%%%%%%%%%%%%%%%%%%%%%%%%%%%%%%%%%%%%
%%%%%%%%%%%%%%%%%%%%%%%%%%%%%%%%%%%%%%%%%%%%%%%%%%%%%%%%%%%%%%%%%%%%%%%%%%

Our work on the definition and the characterization via eqns (\ref{tPhi}) and (\ref{tPhiBis}) of the chronoprojective group, $\Chr(\bR^n)$, of conformal non-relativistic symmetries, as well as its illustration through various examples of spacetime models has shown the richness of the geometric structures involved, namely a mix of Galilei and projective structures in the NC framework. This contrasts with the well-known geometric incompatibility (except in the $1$-dimensional case) between projective and metric conformal structures where the symmetry groups are respectively $\SL(n+1,\bR)$ and $\SO(p+1,q+1)$ in dimension $n=p+q$. 

The fundamental difference with the  familiar extended Schr\"odinger group is that the latter has fixed dynamical exponent $z=2$, and that the central extension subgroup generates (conserved) mass. 
In contrast, the \textit{conformal-NC group} \cite{Duv,BDP0,BDP} has no central extension; however space and time dilations are independent:  there is \emph{no} fixed dynamical exponent,~$z$. 
It is worth mentioning that the recently considered conformal Galilei groups~\cite{DH-NC} allow for any $z=2/N$, where $N=1, 2,\dots$. The case $z=1$ can actually be obtained by contraction from the relativistic conformal group \cite{Ba,Nie3}.

This survey has also helped us reveal the power and usefulness of the Eisenhart-lift to promote these NC symmetries to the realm of Lorentzian conformal geometry through the definition of Bargmann extended spacetime structures as generalized pp-waves. Such a procedure has therefore provided a natural geometric framework for the conformal-Bargmann extensions of the NC spacetime conformal symmetry groups we started with.

Let us finish  by recalling two extra examples where these non-relativistic conformal groups occur in a specific manner. The prototype of conformally-NC flat spacetime is $N\cong(\bR^d\times{}S^1)/\bZ_2$; this M\"obius manifold is in fact a Newton-Hooke spacetime with cosmological constant $\Lambda=d$; cf. \cite{Duv,Duv2,DH}. Also does the symmetry group of the Schr\"odinger-Newton equation \cite{Dio} appear as a subgroup of the conformal-Bargmann group $\Chr(\bR^{d+2})$ featuring a specific dynamical exponent $z=\frac{1}{3}(d+2)$, as shown in \cite{DL} and in agreement with \cite{SZ}.

\bigskip

%%%%%%%%%%%%%%%%%%%%%%%%%%%%%%%%%%%%%%%%%%%%%%%%%%%%%%%%%%%%%%%%%%%%%%%%%%
%%%%%%%%%%%%%%%%%%%%%%%%%%%%%%%%%%%%%%%%%%%%%%%%%%%%%%%%%%%%%%%%%%%%%%%%%%
\textbf{Acknowledgments:}
Thanks are due to M. Cariglia for his constant interest and for helpful suggestions. GWG is grateful to the \textit{Laboratoire de Math\'ematiques et de Physique Th\'eorique de l'Universit\'e de Tours} for hospitality and the R\'egion Centre for a ``Le Studium'' research professorship.
PAH acknowledges partial support from the Chinese Academy of Sciences' Presidential
International Fellowship (Grant No. 2010T1J06), and the \textit{Institute of Modern Physics} of the CAS at Lanzhou for hospitality.
%%%%%%%%%%%%%%%%%%%%%%%%%%%%%%%%%%%%%%%%%%%%%%%%%%%%%%%%%%%%%%%%%%%%%%%%%%
%%%%%%%%%%%%%%%%%%%%%%%%%%%%%%%%%%%%%%%%%%%%%%%%%%%%%%%%%%%%%%%%%%%%%%%%%%

%\phantomsection
\addcontentsline{toc}{section}{References}

\end{document}